\documentclass[conference, a4paper,10pt]{IEEEtran}
\IEEEoverridecommandlockouts

\newcommand{\arxiv}[1]{#1}

\usepackage{amsmath,amssymb,amsfonts}
\usepackage{algorithmic}

\usepackage{xifthen}
\newcommand{\prob}[1][]{%
\ifthenelse{\isempty{#1}}%
      {\ensuremath{P}}%
    {\ensuremath{P\left\(#1\right\)}}%
}

\usepackage{amsthm}
\theoremstyle{remark}

\usepackage{graphicx} %
\graphicspath{{./figures}}
\usepackage{color}
\usepackage[dvipsnames,svgnames, table]{xcolor} %
\usepackage[hidelinks]{hyperref} %
\usepackage{siunitx}
\DeclareSIUnit{\dBm}{dBm}	                %
\DeclareSIUnit{\dBi}{dBi}                   %
\DeclareSIUnit{\dBsm}{dBsm}                 %
\usepackage{booktabs}
\usepackage[backend=biber, style=ieee, citestyle=numeric-comp, maxcitenames=1,mincitenames=1, maxbibnames=1, minbibnames=1,isbn=false, doi=false,citecounter=true]{biblatex}

\usepackage{pgfplots}
\usepgfplotslibrary{groupplots,dateplot}
\usetikzlibrary{patterns,shapes.arrows}
\usetikzlibrary{spy}
\pgfplotsset{compat=newest}
\usetikzlibrary{external}
\usetikzlibrary{calc}
\usetikzlibrary{shadings}
\usetikzlibrary{shadows.blur}
\usetikzlibrary{decorations.pathreplacing}

\usepackage{pgfplotstable}
\pgfplotsset{compat=1.18}

\usepackage{soul}

\pgfmathdeclarefunction{viridis}{1}{%
  \pgfmathparse{%
    1000*(
    0.370489*pow(#1, 3)-
    1.11384*pow(#1, 2)+
    0.929777*(#1)
    )%
  }%
}

\pgfplotsset{every axis/.append style={
    label style={font=\footnotesize},
    legend style={font=\footnotesize},
    tick label style={font=\footnotesize},
}}

\usepackage{tabularx}
\usepackage{tablefootnote}

\AtBeginBibliography{\footnotesize}

\usepackage[font=footnotesize]{caption}
\usepackage{subcaption}
\usepackage[capitalize]{cleveref}
\usepackage{xspace}

\usepackage{placeins}
\usepackage[xindy, nonumberlist]{glossaries}
\makenoidxglossaries%

\newacronym{2d}{2D}{two-dimensional}
\newacronym{3d}{3D}{three-dimensional}
\newacronym{3gpp}{3GPP}{3rd generation partnership project}
\newacronym{4g}{4G}{fourth-generation}
\newacronym{5g}{5G}{fifth-generation}
\newacronym{6dof}{6DoF}{six degrees of freedom}
\newacronym{6g}{6G}{sixth-generation}
\newacronym{abp}{ABP}{authentication by personalisation}
\newacronym{ac}{AC}{alternating current}
\newacronym{aclr}{ACLR}{adjacent channel leakage ratio}
\newacronym{adc}{ADC}{analog-to-digital converter}
\newacronym{adr}{ADR}{adaptive data rate}
\newacronym{aes}{AES}{Advanced Encryption Standard}
\newacronym{afe}{AFE}{analog front end}
\newacronym{ag}{AG}{array gain}
\newacronym{agps}{A-GPS}{Assisted Global Positioning System} %
\newacronym{ai}{AI}{artificial intelligence}
\newacronym{alk}{ALK}{Alkaline}
\newacronym{am}{AM}{amplitude modulation}
\newacronym{amam}{AM/AM}{amplitude modulation to amplitude modulation}
\newacronym{amc}{AMC}{automatic modulation classification}
\newacronym{amp}{AMP}{approximate message passing}
\newacronym{ampm}{AM/PM}{amplitude modulation to phase modulation}
\newacronym{aoa}{AOA}{angle-of-arrival}
\newacronym{aod}{AOD}{angle-of-departure}
\newacronym{ap}{AP}{access point}
\newacronym{api}{API}{application program interface}
\newacronym{apt}{APT}{acoustic power transfer}
\newacronym{apu}{APU}{access point unit}
\newacronym{ar}{AR}{augmented reality}
\newacronym{arima}{ARIMA}{Auto-Regressive Integrated Moving Average}
\newacronym{arp}{ARP}{Antenna Reference Point}
\newacronym{asic}{ASIC}{application specific integrated circuit}
\newacronym{ask}{ASK}{amplitude-shift keying}
\newacronym{at}{AT}{ATtention}
\newacronym{auv}{AUV}{autonomous underwater vehicle}
\newacronym{awg}{AWG}{arbitrary waveform generator}
\newacronym{awgn}{AWGN}{additive white Gaussian noise}
\newacronym{baw}{BAW}{bulk acoustic wave}
\newacronym{bb}{BB}{base-band}
\newacronym{bc}{BC}{backscatter communication}
\newacronym{bd}{BD}{backscatter device}
\newacronym{be}{BE}{Belgium}
\newacronym{ber}{BER}{bit error rate}
\newacronym{bf}{BF}{beamforming}
\newacronym{bga}{BGA}{ball grid array}
\newacronym{bldc}{BLDC}{brushless DC}
\newacronym{bod}{BOD}{Brown-Out Detection}
\newacronym{bom}{BOM}{bill of materials}
\newacronym{bpsk}{BPSK}{binary phase-shift keying}
\newacronym{bs}{BS}{base station}
\newacronym{bu}{BU}{booster unit}
\newacronym{bw}{BW}{bandwidth}
\newacronym{ca}{CA}{carrier emitter}
\newacronym{cad}{CAD}{channel activity detection}
\newacronym{cars}{CARS}{calibration reference signal}
\newacronym{cbm}{CBM}{condition based maintenance}
\newacronym{cc}{CC}{constant current}
\newacronym{ccdf}{CCDF}{complementary cumulative distribution function}
\newacronym{ccnn}{CCNN}{circular convolutional neural network}
\newacronym{ccs}{CCS}{correlative channel sounder}
\newacronym{cdf}{CDF}{cumulative distribution function}
\newacronym{cdma}{CDMA}{code division-multiple access}
\newacronym{cdrx}{CDRX}{connected mode DRX}
\newacronym{ce}{CE}{coverage enhancement}
\newacronym{ced}{CED}{cumulative energy density}
\newacronym{cf}{CF}{cell-free}
\newacronym{cfmmimo}{CF-mMIMO}{cell-free massive MIMO}
\newacronym{cfo}{CFO}{carrier frequency offset}
\newacronym{cir}{CIR}{channel impulse response}
\newacronym{cla}{CLA}{closed-loop approach}
\newacronym{clk}{CLK}{clock}
\newacronym{cmos}{CMOS}{complementary metal oxide semiconductor}
\newacronym{cost}{COST}{commercial off-the-shelf}
\newacronym{cots}{COTS}{commercial off-the-shelf}
\newacronym{cp}{CP}{cyclic prefix}
\newacronym{cpt}{CPT}{capacitive power transfer}
\newacronym{cpu}{CPU}{central-processing unit}
\newacronym{cpw}{CPW}{coplanar waveguide}
\newacronym{cqi}{CQI}{channel quality indicator}
\newacronym{cr}{CR}{coding rate}
\newacronym{crc}{CRC}{cyclic redundancy check}
\newacronym{crlb}{CRLB}{Cram\'er-Rao lower bound}
\newacronym{crs}{CRS}{cell reference signal}
\newacronym{cs}{CS}{compressed sensing}
\newacronym{cse}{CSE}{channel state estimation}
\newacronym{csi}{CSI}{channel state information}
\newacronym{csp}{CSP}{contact service point}
\newacronym{css}{CSS}{chirp spread spectrum}
\newacronym{cu}{CU}{central unit}
\newacronym{cv}{CV}{constant voltage}
\newacronym{cw}{CW}{continuous wave}
\newacronym{d2d}{D2D}{device-to-device}
\newacronym{dab}{DAB}{Digital Audio Broadcasting}
\newacronym{dac}{DAC}{digital-to-analog converter}
\newacronym{daq}{DAQ}{data acquisition system}
\newacronym{das}{DAS}{distributed antenna systems}
\newacronym{dbpsk}{DBPSK}{Differential Binary Phase Shift Keying}
\newacronym{dc}{DC}{direct current}
\newacronym{dcc}{DCC}{dynamic cooperation clustering}
\newacronym{ddc}{DDC}{digital down conversion}
\newacronym{de}{DE}{drain efficiency}
\newacronym{dft}{DFT}{discrete Fourier transform}
\newacronym{dl}{DL}{downlink}
\newacronym{dlc}{DLC}{Distributed Laser Charging}
\newacronym{dli}{DLI}{direct link interference}
\newacronym{dlt}{DLT}{Distributed Ledger Technology}
\newacronym{dm}{DM}{diffuse multipath}
\newacronym{dma}{DMA}{Direct Memory Access}
\newacronym{dmac}{DMAC}{Direct Memory Access Controller}
\newacronym{dmc}{DMC}{diffuse multipath component}
\newacronym{dmimo}{D-MIMO}{distributed MIMO}
\newacronym{doa}{DOA}{direction-of-arrival}
\newacronym{dp}{DP}{differencial privacy}
\newacronym{dpd}{DPD}{digital pre-distortion}
\newacronym{dpdk}{DPDK}{Data Plane Development Kit}
\newacronym{dram}{DRAM}{dynamic random-access memory}
\newacronym{drcs}{$\Delta$RCS}{differential-radar cross section}
\newacronym{drx}{DRX}{Discontinuous Reception Mode}
\newacronym{dsb}{DSB}{double-sideband}
\newacronym{dsl}{DSL}{digital subscriber line}
\newacronym{dsp}{DSP}{digital signal processing}
\newacronym{dss}{DSS}{Dataset Storage Standard}
\newacronym{duc}{DUC}{digital up-converter}
\newacronym{dvb}{DVB}{Digital Video Broadcasting}
\newacronym{e2e}{E2E}{end-to-end}
\newacronym{easa}{EASA}{European Union Aviation Safety Agency}
\newacronym{ebg}{EBG}{electromagnetic bandgap}
\newacronym{ec}{EC}{European Commission}
\newacronym{ecc}{ECC}{elliptic curve cryptography}
\newacronym{ecdf}{eCDF}{empirical cumulative distribution function}
\newacronym{ecdlp}{ECDLP}{elliptic curve discrete logarithm problem}
\newacronym{ecsp}{ECSP}{edge computing service point}
\newacronym{edlc}{EDLC}{electrostatic double-layer capacitors}
\newacronym{edrx}{eDRX}{Extended Discontinuous Reception Mode}
\newacronym{ee}{EE}{energy efficiency}
\newacronym{egprs}{EGPRS}{Enhanced Data Rates for GSM Evolution}
\newacronym{eh}{EH}{energy harvesting}
\newacronym{eirp}{EIRP}{equivalent isotropically radiated power}
\newacronym{em}{EM}{electromagnetic}
\newacronym{embb}{eMBB}{enhanced Mobile Broadband}
\newacronym{en}{EN}{energy neutral}
\newacronym{end}{END}{energy-neutral device}
\newacronym{enob}{ENOB}{effective number of bits}
\newacronym{eol}{EoL}{end of life}
\newacronym{epu}{EPU}{edge processing unit}
\newacronym{er}{ER}{Energy Receiver}
\newacronym{erp}{ERP}{effective radiation power}
\newacronym[plural=ESCs,firstplural=electronic speed controllers (ESCs)]{esc}{ESC}{electronic speed control}
\newacronym{esd}{ESD}{electrostatic discharge}
\newacronym{esl}{ESL}{electronic shelf label}
\newacronym{esr}{ESR}{equivalent series resistance}
\newacronym{et}{ET}{Energy Transmitter}
\newacronym{etsi}{ETSI}{European Telecommunications Standards Institute}
\newacronym{evd}{EVD}{eigenvalue decomposition}
\newacronym{evm}{EVM}{Error Vector Magnitude}
\newacronym{ewlb}{eWLB}{Embedded Wafer Level Ball Grid Array}
\newacronym{fa}{FA}{federation anchor}
\newacronym{fair}{FAIR}{Findability, Accessibility, Interoperability, and Reuse of digital assets}
\newacronym{fd}{FD}{front-haul distance}
\newacronym{fdd}{FDD}{frequency-division duplexing}
\newacronym{fde}{FDE}{frequency domain equalizer}
\newacronym{fdm}{FDM}{frequency-division multiplexing}
\newacronym{fdma}{FDMA}{frequency division multiple access}
\newacronym{fem}{FEM}{finite element analysis}
\newacronym{fembb}{feMBB}{further enhanced mobile broadband}
\newacronym{fft}{FFT}{fast Fourier transform}
\newacronym{fh}{FH}{fronthaul}
\newacronym{fhss}{FHSS}{frequency hopping spread spectrum}
\newacronym{fifo}{FIFO}{First In, First Out}
\newacronym{fim}{FIM}{Fisher information matrix}
\newacronym{fits}{FITS}{Flexible Image Transport System}
\newacronym{fl}{FL}{federated learning}
\newacronym{fom}{FoM}{figure of merit}
\newacronym{fpga}{FPGA}{field-programmable gate array}
\newacronym{fram}{FRAM}{Ferroelectric Random Access Memory}
\newacronym{fsk}{FSK}{frequency shift keying}
\newacronym{fss}{FSS}{frequency selective surface}
\newacronym{gb}{GB}{grant-based}
\newacronym{gdpr}{GDPR}{general data protection regulation}
\newacronym{gf}{GF}{grant-free}
\newacronym{gmsk}{GMSK}{Gaussian minimum-shift keying}
\newacronym{gnb}{gNB}{Next Generation Node B}
\newacronym{gni}{GNI}{gross national income}
\newacronym{gnn}{GNN}{graph neural network}
\newacronym{gnss}{GNSS}{global navigation satellite system}
\newacronym{gpclk}{GPCLK}{general purpose clock}
\newacronym{gpio}{GPIO}{General-Purpose Input/Output}
\newacronym{gpl}{GPL}{GNU General Public License}
\newacronym{gprs}{GPRS}{General Packet Radio Services}
\newacronym{gps}{GPS}{Global Positioning System}
\newacronym{gpu}{GPU}{graphical processing unit}
\newacronym{grc}{GRC}{GNU Radio Companion}
\newacronym{gscm}{GSCM}{geometry‐based stochastic model}
\newacronym{gsm}{GSM}{Global System for Mobile Communications}  %
\newacronym{gwp}{GWP}{Global Warming Potential}
\newacronym{harq}{HARQ}{hybrid automatic repeat request}
\newacronym{hat}{HAT}{hardware attached on top}
\newacronym{hcs}{HCS}{human-centric services}
\newacronym{hdf5}{HDF5}{Hierarchical Data Format version 5}
\newacronym{hfss}{HFSS}{High Frequency Simulator Software}
\newacronym{hmd}{HMD}{head-mounted display}
\newacronym{hpbm}{HPBM}{half power beam width}
\newacronym{HyMPRo}{HyMPRo}{Hybrid Multi-Path Routing algorithm}
\newacronym{i.i.d.}{i.i.d.}{independent and identically distributed}
\newacronym{i2c}{I2C}{Inter-Integrated Circuit}
\newacronym{iaq}{IAQ}{Indoor Air Quality}
\newacronym{ib}{IB}{in-band}
\newacronym{ibo}{IBO}{input back-off}
\newacronym{ic}{IC}{integrated circuit}
\newacronym{ici}{ICI}{intercarrier interference}
\newacronym{icnirp}{ICNIRP}{International Commission on Non-Ionizing Radiation Protection}
\newacronym{id}{ID}{information decoding}
\newacronym{idft}{IDFT}{inverse discrete Fourier transform}
\newacronym{if}{IF}{intermediate-frequency}
\newacronym{iid}{i.i.d.}{independently and identically distributed}
\newacronym{iis}{IIS}{integrated information system}
\newacronym{im}{IM}{intermodulation}
\newacronym{imd}{IMD}{intermodulation distortion}
\newacronym{imu}{IMU}{inertial measurement unit}
\newacronym{io}{IO}{input/output}
\newacronym{ioe}{IoE}{Internet of Everything}
\newacronym{iot}{IoT}{Internet of Things}
\newacronym{ipt}{IPT}{inductive power transfer}
\newacronym{ipy}{IPY}{Interventions per Year}
\newacronym{iq}{IQ}{in-phase and quadrature}
\newacronym{iqi}{IQI}{IQ imbalance}
\newacronym{ir}{IR}{infrared}
\newacronym{isi}{ISI}{intersymbol interference}
\newacronym{ism}{ISM}{industrial, scientific and medical}
\newacronym{isp}{ISP}{internet service provider}
\newacronym{jesd}{JESD}{Joint Electron Devices Engineering Council}
\newacronym{jfet}{JFET}{junction field effect transistor}
\newacronym{kpi}{KPI}{key performance indicator}
\newacronym{kvi}{KVI}{key value indicator}
\newacronym{larva}{LARVA}{LARge Virtual Array}
\newacronym{lca}{LCA}{life cycle assessment}
\newacronym{lco}{LCO}{lithium cobalt oxide}
\newacronym{ldo}{LDO}{Low-dropout voltage regulator}
\newacronym{ldpc}{LDPC}{low-density parity-check}
\newacronym{led}{LED}{Light Emitting Diode}
\newacronym{less}{LESS}{Low Energy Scheduler Solution}
\newacronym{lfp}{LFP}{lithium iron phosphate}
\newacronym{lib}{LIB}{Lithium-Ion Battery}
\newacronym{lic}{LIC}{lithium-ion capacitor}
\newacronym{lid}{LID}{Lithium Iron Disulfide}
\newacronym{lidar}{LiDAR}{light detection and ranging}
\newacronym{liion}{Li-ion}{lithium-ion}
\newacronym{lipo}{LiPo}{lithium polymer}
\newacronym{lis}{LIS}{large intelligent surface}
\newacronym{llh}{LLH}{log-likelihood}
\newacronym{lmd}{LMD}{Lithium Manganese Dioxide}
\newacronym{lmmse}{LMMSE}{least minimum mean square error}
\newacronym{lmo}{LMO}{lithium ion manganese oxide}
\newacronym{lna}{LNA}{low-noise amplifier}
\newacronym{lo}{LO}{local oscillator}
\newacronym{lora}{LoRa}{long range}
\newacronym{lorawan}{LoRaWAN}{long-range wide-area network}
\newacronym{los}{LoS}{line-of-sight}
\newacronym{lp}{LP}{linear programming}
\newacronym{lpf}{LPF}{low-pass filter}
\newacronym{lpt}{LPT}{laser power transfer}
\newacronym{lpwa}{LPWA}{Low Power Wide Area}
\newacronym{lpwan}{LPWAN}{low-power wide-area network}
\newacronym{lpwans}{LPWANs}{Low-Power Wide-Area Networks}
\newacronym{lqi}{LQI}{link quality indicator}
\newacronym{lrelu}{LReLU}{leaky rectified linear unit}
\newacronym{lrt}{LRT}{likelihood-ratio test}
\newacronym{ls}{LS}{least squares}
\newacronym{lsa}{LSA}{large synthetic array}
\newacronym{lsf}{LSF}{large-scale fading}
\newacronym{lsfc}{LSFC}{large-scale fading component}
\newacronym{lstm}{LSTM}{Long Short-Term Memory}
\newacronym{ltc}{LTC}{lithium thionyl chloride}
\newacronym{lte}{LTE}{Long Term Evolution}
\newacronym{lti}{LTI}{linear time-invariant}
\newacronym{lto}{LTO}{lithium titanate}
\newacronym{lusta}{LUSTA 5G }{Logistique mUltimodale Sécuritaire Téléopérée \& Autonome 5G}
\newacronym{m2m}{M2M}{machine to machine}
\newacronym{mac}{MAC}{Medium Access Control}
\newacronym{mate}{MATE}{millimeter-wave MIMO testbed}
\newacronym{mc}{MC}{Monte Carlo}
\newacronym{mcl}{MCL}{Maximum Coupling Loss}
\newacronym{mcs}{MCS}{modulation and coding scheme}
\newacronym{mcu}{MCU}{microcontroller unit}
\newacronym{mec}{MEC}{multi-access edge computing}
\newacronym{mems}{MEMS}{micro-electromechanical systems}
\newacronym{mf}{MF}{matched filter}
\newacronym{mimo}{MIMO}{multiple-input multiple-output}
\newacronym{miso}{MISO}{multiple-input single-output}
\newacronym{ml}{ML}{machine learning}
\newacronym{mlp}{MLP}{multilayer perceptron}
\newacronym{mmic}{MMIC}{monolithic microwave integrated circuit}
\newacronym{mmimo}{mMIMO}{massive MIMO}
\newacronym{mmse}{MMSE}{minimum mean square error}
\newacronym{mmtc}{mMTC}{massive machine-typed communication}
\newacronym{mmwave}{mmWave}{millimeter wave}
\newacronym{mn}{MN}{matching network}
\newacronym{mosfet}{MOSFET}{metal-oxide semiconductor field effect transistor}
\newacronym{mpc}{MPC}{multipath component}
\newacronym{mppt}{MPPT}{maximum power point tracking}
\newacronym{mr}{MR}{maximum ratio}
\newacronym{mrc}{MRC}{maximum ratio combining}
\newacronym{mrc_em}{MRC}{maximum ratio combining}
\newacronym{mrc_EM}{MRC}{Magnetic Resonance Coupling}
\newacronym{mrt}{MRT}{maximum ratio transmission}
\newacronym{mse}{MSE}{mean square error}
\newacronym{msk}{MSK}{Minimum-Shift Keying}
\newacronym{mtc}{MTC}{Machine-Type Communication}
\newacronym{multi-rat}{Multi-RAT}{multiple radio access technology}
\newacronym{multirat}{Multi-RAT}{Multiple Radio Access Technology}
\newacronym{music}{MUSIC}{MUltiple SIgnal Classification}
\newacronym{navauwall}{NAVAUWALL}{AUtomated NAVigation in WALLonia}
\newacronym{nb}{NB}{narrowband}
\newacronym{nbiot}{NB-IoT}{narrowband IoT}
\newacronym{nca}{NCA}{nickel cobalt aluminum}
\newacronym{netcdf}{NetCDF}{Network Common Data Form}
\newacronym{nf}{NF}{noise figure}
\newacronym{nfc}{NFC}{near-field communication}
\newacronym{nfv}{NFV}{network function virtualization}
\newacronym{ngmn}{NGMN}{Next Generation Mobile Networks }
\newacronym{ni}{NI}{National Instruments}
\newacronym{nicd}{NiCd}{nikkel cadmium}
\newacronym{nimh}{NiMH}{nikkel metal hydride}
\newacronym{nlos}{NLoS}{non-line-of-sight}
\newacronym{nmc}{NMC}{nickel manganese cobalt}
\newacronym{nmos}{nMOS}{n-channel metal-oxide semiconductor}
\newacronym{nn}{NN}{neural network}
\newacronym{nnls}{NNLS}{non-negative least squares}
\newacronym{noma}{NOMA}{non-orthogonal multiple access}
\newacronym{np}{NP}{Neyman-Pearson}
\newacronym{npbch}{NPBCH}{Narrowband Physical Broadcast Channel}
\newacronym{npss}{NPSS}{Narrow Band Primay Synchronization Signal}
\newacronym{nr}{NR}{New Radio}
\newacronym{nrs}{NRS}{Narrow Band Reference Signal}
\newacronym{nsss}{NSSS}{Narrowband Secondary Synchronization Signal}
\newacronym{ntp}{NTP}{network time protocol}
\newacronym{oai}{OAI}{OpenAirInterface} %
\newacronym{obw}{OBW}{occupied bandwidth}
\newacronym{ofdm}{OFDM}{orthogonal frequency-division multiplexing}
\newacronym{ofdma}{OFDMA}{orthogonal frequency-division multiple access}
\newacronym{ofdmim}{OFDM-IM}{OFDM with index modulation}
\newacronym{oma}{OMA}{orthogonal multiple access}
\newacronym{oob}{OOB}{out-of-band}
\newacronym{ook}{OOK}{on-off keying}
\newacronym{oran}{O-RAN}{open radio-access network}
\newacronym{os}{OS}{operating system}
\newacronym{ota}{OTA}{over-the-air}
\newacronym{otaa}{OTAA}{over-the-air authentication}
\newacronym{p1}{P1}{Phase 1}
\newacronym{p2}{P2}{Phase 2}
\newacronym{p2p}{P2P}{point-to-point}
\newacronym{pa}{PA}{power amplifier}
\newacronym{pae}{PAE}{power-added efficiency}
\newacronym{pana}{PanA}{Panel A}
\newacronym{panb}{PanB}{Panel B}
\newacronym{papr}{PAPR}{peak-to-average power ratio}
\newacronym{pc}{PC}{pilot count}
\newacronym{pcb}{PCB}{printed circuit board}
\newacronym{pcg}{PCG}{power consumption gain}
\newacronym{pcie}{PCIe}{Peripheral Component Interconnect Express}
\newacronym{pcsi}{PCSI}{perfect channel state information}
\newacronym{pd}{PD}{powered device}
\newacronym{pdcch}{PDCCH}{physical downlink control channel}
\newacronym{pdf}{PDF}{probability density function}
\newacronym{pdp}{PDP}{power delay profile}
\newacronym{pdsch}{PDSCH}{physical downlink shared channel}
\newacronym{pe}{PE}{processing element}
\newacronym{peb}{PEB}{positioning error bound}
\newacronym{per}{PER}{packet error rate}
\newacronym{pet}{PET}{privacy enhancing technology}
\newacronym{pg}{PG}{path gain}
\newacronym{pgd}{PGD}{proximal gradient descent}
\newacronym{phy}{PHY}{physical}
\newacronym{pki}{PKI}{public key infrastucture}
\newacronym{pl}{PL}{path loss}
\newacronym{pll}{PLL}{phase-locked loop}
\newacronym{pmf}{PMF}{polymer microwave fiber}
\newacronym{pmu}{PMU}{Power Management Unit}
\newacronym{pn}{PN}{pseudo-noise}
\newacronym{poe}{PoE}{power-over-Ethernet}
\newacronym{pps}{1PPS}{pulse per second}
\newacronym{pr}{PR}{phase reversal}
\newacronym{prbs}{PRBs}{Physical Resource Blocks}
\newacronym{prs}{PRS}{Peripheral Reflex System}
\newacronym{ps}{PS}{Processing System}
\newacronym{psd}{PSD}{power spectral density}
\newacronym{pse}{PSE}{power sourcing equipment}
\newacronym{psk}{PSK}{phase shift keying}
\newacronym{psm}{PSM}{power saving mode}
\newacronym{pss}{PSS}{primary synchronisation signal}
\newacronym{ptp}{PTP}{precision-time protocol}
\newacronym{ptrs}{PTRS}{Phase-Tracking Reference Signals}
\newacronym{ptw}{PTW}{paging time window}
\newacronym{pv}{PV}{photovoltaic}
\newacronym{pw}{PW}{planar wavefront}
\newacronym{pwm}{PWM}{pulse width modulation}
\newacronym{qam}{QAM}{quadrature amplitude modulation}
\newacronym{qos}{QoS}{quality-of-service}
\newacronym{qpsk}{QPSK}{quadrature phase-shift keying}
\newacronym{qrrls}{QR-RLS}{QR decomposition based recursive least squares}
\newacronym{quadriga}{QuaDRiGa}{QUAsi Deterministic RadIo channel GenerAtor}
\newacronym{ra}{RA}{Random Access}
\newacronym{ram}{RAM}{random-access memory}
\newacronym{ran}{RAN}{radio access network}
\newacronym{rar}{RAR}{Random Access Response}
\newacronym{rat}{RAT}{radio access technology}
\newacronym{rb}{Rb}{Rubidium}
\newacronym{rbs}{RBS}{radio base station}
\newacronym{rbw}{RBW}{resolution bandwidth}
\newacronym{rcs}{RCS}{radar cross section}
\newacronym{rdl}{RDL}{redistribution layer}
\newacronym{re}{RE}{radio element}
\newacronym{relu}{ReLU}{rectified linear unit}
\newacronym{rf}{RF}{radio frequency}
\newacronym{rfeh}{RFEH}{radio frequency energy harvesting}
\newacronym{rfic}{RFIC}{radio-frequency integrated circuit}
\newacronym{rfid}{RFID}{radio frequency identification}
\newacronym{rfpt}{RFPT}{radio frequency power transfer}
\newacronym{rfsoc}{RFSoC}{Radio Frequency System-on-Chip}
\newacronym{rir}{RIR}{room impulse response}
\newacronym{ris}{RIS}{reflective intelligent surface}%
\newacronym{rllmtc}{RLLMTC}{reliable low latency machine type communication}
\newacronym{rls}{RLS}{recursive least squares}
\newacronym{rms}{RMS}{root-mean-square}
\newacronym{rmse}{RMSE}{root-mean-square error}
\newacronym{rmt}{RMT}{random matrix theory}
\newacronym{rof}{RoF}{radio-over-fiber}
\newacronym{ros}{ROS}{robot operating system}
\newacronym{rpi}{RPi}{Raspberry Pi}
\newacronym{rrc}{RRC}{Radio Resource Connection}
\newacronym{rreq}{RREQ}{route request packet}
\newacronym{rsrp}{RSRP}{Reference Signals Received Power}
\newacronym{rsrq}{RSRQ}{Reference Signal Received Quality}
\newacronym{rss}{RSS}{received signal strength}
\newacronym{rssi}{RSSI}{received signal strength indicator}
\newacronym{rtc}{RTC}{real time clock}
\newacronym{rtf}{RTF}{reader talks first}
\newacronym{rtk}{RTK}{real time kinematics}
\newacronym{rts}{RTS}{ray tracing simulator}
\newacronym{ru}{RU}{Radio Unit}
\newacronym{rv}{RV}{random variable}
\newacronym{rw}{RW}{RadioWeaves}
\newacronym{rx}{RX}{receiver}
\newacronym{rzf}{RZF}{regularized zero forcing}
\newacronym{s-parameter}{S-parameter}{scattering parameter}
\newacronym{sa}{SA}{synchronization anchor}
\newacronym{sa5}{SA}{Stand Alone}
\newacronym{sar}{SAR}{specific absorption rate}
\newacronym{sbl}{SBL}{sparse Bayesian learning}
\newacronym{scfdma}{SCFDMA}{single-carrier frequency division multiple access}
\newacronym{sdg}{SDG}{Sustainable Development Goal}
\newacronym{sdm}{SDM}{sigma-delta modulator}
\newacronym{sdma}{SDMA}{spatial-division multiple access}
\newacronym{sdn}{SDN}{software-defined network}
\newacronym{sdof}{SDoF}{sigma-delta over fiber}
\newacronym{sdr}{SDR}{software-defined radio}
\newacronym{se}{SE}{spectral efficiency}
\newacronym{sei}{SEI}{specific emitter identification}
\newacronym{ser}{SER}{symbol-error rate}
\newacronym{sf}{SF}{spreading factor}
\newacronym{sfn}{SFN}{single frequency network}
\newacronym{sfp}{SFP}{small form-factor pluggable}
\newacronym{sha}{SHA}{Secure Hash Algorithm}
\newacronym{SigMF}{SigMF}{Signal Metadata Format}
\newacronym{simo}{SIMO}{single-input multiple-output}
\newacronym{sinr}{SINR}{signal-to-interference-plus-noise ratio}
\newacronym{siso}{SISO}{single-input single-output}
\newacronym{slam}{SLAM}{simultaneous localization and mapping}
\newacronym{slc}{SLC}{spatial leakage suppression}
\newacronym{slerp}{SLERP}{spherical linear interpolation}
\newacronym{sma}{SMA}{SubMiniature version A}
\newacronym{smc}{SMC}{specular multipath component}
\newacronym{smps}{SMPS}{switched mode power supply}
\newacronym{sndr}{SNDR}{signal-to-noise-and-distortion ratio}
\newacronym{snidr}{SNIDR}{signal-to-noise-and-interference-and-distortion ratio}
\newacronym{snir}{SNIR}{signal-to-interference-plus-noise ratio}
\newacronym{snr}{SNR}{signal-to-noise ratio}
\newacronym{soc}{SoC}{state of charge}
\newacronym{SoC}{SOC}{System on Chip}
\newacronym{sp}{SP}{service point}
\newacronym{spdt}{SPDT}{single pole double throw}
\newacronym{spi}{SPI}{Serial Peripheral Interface}
\newacronym{spst}{SPST}{single pole single throw}
\newacronym{sram}{SRAM}{static random-access memory}
\newacronym{srd}{SRD}{short-range device}
\newacronym{srls}{SRLS}{standard recursive least squares}
\newacronym{srs}{SRS}{Sounding Reference Signal}
\newacronym{ssb}{SSB}{synchronisation signal block}
\newacronym{ssd}{SSD}{solid state drive}
\newacronym{ssq}{SSQ}{simulator sickness questionnaire}
\newacronym{steam}{STEAM}{science, technology, engineering, the arts, and mathematics}
\newacronym{svd}{SVD}{singular value decomposition}
\newacronym{sw}{SW}{spherical wavefront}
\newacronym{swipt}{SWIPT}{simultaneous wireless information and power transfer}
\newacronym{synce}{SyncE}{Synchronous Ethernet}
\newacronym{tau}{TAU}{tracking area update}
\newacronym{tcer}{TCER}{transported to consumed energy ratio}
\newacronym{tcp}{TCP}{Transmission Control Protocol}
\newacronym{tcxo}{TCXO}{temperature compensated crystal oscillator}
\newacronym{tdd}{TDD}{time division duplexing}
\newacronym{tdma}{TDMA}{time division-multiple access}
\newacronym{tdoa}{TDOA}{time-difference-of-arrival}
\newacronym{thz}{THz}{Terahertz}
\newacronym{to}{TO}{timing offset}
\newacronym{toa}{TOA}{time-of-arrival}
\newacronym{tof}{ToF}{time-of-flight}
\newacronym{tosm}{TOSM}{through-open-short-match}
\newacronym{tpms}{TPMS}{Tire-Pressure Monitoring System}
\newacronym{trl}{TRL}{technology readyness level}
\newacronym{trp}{TRP}{Transmission Reception Point}
\newacronym{tsn}{TSN}{time-sensitive networking}
\newacronym{ttf}{TTF}{tag talks first}
\newacronym{ttff}{TTFF}{Time To First Fix}
\newacronym{ttm}{TTM}{time to market}
\newacronym{ttn}{TTN}{The Things Network}
\newacronym{tx}{TX}{transmitter}
\newacronym{uart}{UART}{Universal Asynchronous Receiver/Transmitter}
\newacronym{uav}{UAV}{unmanned aerial vehicle}
\newacronym{uc}{UC}{use case}
\newacronym{ucie}{UCIe}{Universal Chiplet Interconnect Express}
\newacronym{udp}{UDP}{User Datagram Protocol}
\newacronym{ue}{UE}{user equipment}
\newacronym{ugv}{UGV}{unmanned ground vehicle}
\newacronym{uhd}{UHD}{USRP hardware driver}
\newacronym{uhf}{UHF}{ultra-high frequency}
\newacronym{ul}{UL}{uplink}
\newacronym{ula}{ULA}{uniform linear array}
\newacronym{uMIMO}{$\mu$-MIMO}{ultra-massive Multiple-Input Multiple-Output}
\newacronym{ummtc}{umMTC}{ultra massive machine type communication}
\newacronym{un}{UN}{United Nations}
\newacronym{upa}{UPA}{uniform planar array}
\newacronym{ura}{URA}{uniform rectangular array}
\newacronym{urllc}{URLLC}{ultra-reliable low-latency communications}
\newacronym{usrp}{USRP}{universal software radio peripheral}
\newacronym{uv}{UV}{unmanned vehicle}
\newacronym{uwb}{UWB}{ultrawideband}
\newacronym{v2v}{V2V}{vehicle-to-vehicle}
\newacronym{vco}{VCO}{voltage-controlled oscillator}
\newacronym{vep}{VEP}{virtual edge platform}
\newacronym{vlc}{VLC}{visible light communication}
\newacronym{vlp}{VLP}{visible light positioning}
\newacronym{vna}{VNA}{vector network analyzer}
\newacronym{voc}{VOC}{Voltatile Organic Compound}
\newacronym{votable}{VOTable}{Virtual Observatory Table}
\newacronym{vr}{VR}{virtual reality}
\newacronym{wb}{WB}{wideband}
\newacronym{wban}{WBAN}{wireless body area network}
\newacronym{wd}{WD}{Wireless distance}
\newacronym{wimax}{WiMAX}{Worldwide Interoperability for Microwave Access}
\newacronym{wlan}{WLAN}{wireless LAN}
\newacronym{wpt}{WPT}{wireless power transfer}
\newacronym{wr}{WR}{White Rabbit}
\newacronym{wrsn}{WRSN}{wireless rechargeable sensor network}
\newacronym{wsn}{WSN}{wireless sensor network}
\newacronym{xets}{XETS}{cross exponentially tapered slot}
\newacronym{xr}{XR}{extended reality}
\newacronym{z3ro}{Z3RO}{zero third-order distortion}
\newacronym{zf}{ZF}{zero-forcing}
\newacronym{zmcscg}{ZMCSCG}{zero mean circularly symmetric complex Gaussian}
\newacronym{ep}{EP}{energy profiler}

\newacronym{cordic}{CORDIC}{coordinate rotation digital computer}

\newacronym{zmq}{ZMQ}{ZeroMQ}%

\usepackage{comment}

\newcolumntype{R}{>{\raggedleft\arraybackslash}X}

\usepackage{graphicx}
\usepackage{textcomp}
\usepackage{xcolor}
\def\BibTeX{{\rm B\kern-.05em{\sc i\kern-.025em b}\kern-.08em
    T\kern-.1667em\lower.7ex\hbox{E}\kern-.125emX}}
 
\newcommand{\tikzxmark}{%
\tikz[scale=0.23] {
    \draw[line width=0.7,line cap=round] (0,0) to [bend left=6] (1,1);
    \draw[line width=0.7,line cap=round] (0.2,0.95) to [bend right=3] (0.8,0.05);
}}
\newcommand{\tikzcmark}{%
\tikz[scale=0.23] {
    \draw[line width=0.7,line cap=round] (0.25,0) to [bend left=10] (1,1);
    \draw[line width=0.8,line cap=round] (0,0.35) to [bend right=1] (0.23,0);
}}
\newcommand{\tikzpmmark}{%
\tikz[scale=0.23] {
    \draw[line width=0.7,line cap=round] (0.5,0.2) to [bend left=6] (0.5,1);
    \draw[line width=0.7,line cap=round] (0.1,0.65) to [bend right=3] (0.9,0.65);
    \draw[line width=0.7,line cap=round] (0.1,0.05) to [bend right=3] (0.9,0.05);
}}

\definecolor{darkgray176}{RGB}{176,176,176}
\definecolor{BF}{RGB}{228,26,28}
\definecolor{AS}{RGB}{152,78,163}
\definecolor{SISO}{RGB}{77,175,74}
\definecolor{G-BF}{RGB}{55,126,184}

\newcommand{\BF}{\textcolor{BF}{BF}\xspace}
\newcommand{\GBF}{\textcolor{G-BF}{G-BF}\xspace}
\newcommand{\AS}{\textcolor{AS}{RPS}\xspace}

\newcommand{\SISO}{\textcolor{SISO}{SISO}\xspace}

 \usepackage{fontawesome}

 \makeatletter
\DeclareRobustCommand\bfseries{%
  \not@math@alphabet\bfseries\mathbf
  \fontseries\bfdefault\selectfont
  \boldmath %
}
\makeatother

\usepackage{enumitem}
\usepackage[dvipsnames,svgnames, table]{xcolor} %

\usepackage{listofitems}
\usepackage{pgffor}

\makeatletter
\def\myColorList{LimeGreen,BrickRed,Fuchsia,Bittersweet,YellowOrange, YellowGreen, WildStrawberry} %
\readlist\ColorList\myColorList

\newcommand{\defineauthors}[1]{

    \foreach \x [count=\xi from 1] in {#1} { 
    \expandafter\xdef\csname\x\endcsname####1{\noexpand\textcolor{\ColorList[\xi]}{[\unexpanded\expandafter{\x}: ####1]}}%
    }
}
\makeatother

\usepackage{array}
\newcolumntype{H}{>{\setbox0=\hbox\bgroup}c<{\egroup}@{}}

\newcounter{assumcounter}
\usepackage{tikz}
\usetikzlibrary{external}
\usepackage{fontawesome}

\tikzset{
    every node/.append style={font=\footnotesize},
    every label/.append style={font=\footnotesize}
}
\pgfplotsset{
    every axis legend/.append style={
        fill opacity=0.8,
        draw=none,
    },
    every axis/.append style={
        x grid style={darkgray!60},
        xmajorgrids,
        y grid style={darkgray!60},
        ymajorgrids,
    }
}

\begin{document}\sloppy
\title{Experimental Study on the Effect of Synchronization Accuracy for Near-Field RF Wireless Power Transfer in Multi-Antenna Systems
\thanks{The work is supported by the REINDEER project under grant agreement No.~101013425 and by the e-construct project under the 'Vlaams Agentschap Innoveren en Ondernemen (VLAIO) grant agreement HBC.2021.0911.}%
\arxiv{\thanks{Manuscript accepted on 12 December 2024 for 2025 19th European Conference on Antennas and Propagation (EuCAP)~\cite{Call2503:Experimental}.}}
}

\author{
    \IEEEauthorblockN{Gilles Callebaut\IEEEauthorrefmark{1}, Jarne Van Mulders\IEEEauthorrefmark{1}, Bert Cox\IEEEauthorrefmark{1}, Benjamin J.\,B. Deutschmann\IEEEauthorrefmark{2}, \\ Geoffrey Ottoy\IEEEauthorrefmark{1}, Lieven De Strycker\IEEEauthorrefmark{1} and Liesbet Van der Perre\IEEEauthorrefmark{1}}
    \IEEEauthorblockA{\IEEEauthorrefmark{1}\textit{KU Leuven,} Belgium, gilles.callebaut@kuleuven.be, \IEEEauthorrefmark{2}\textit{Graz University of Technology,} Austria}
}

\maketitle

\begin{abstract}
Wireless power transfer (WPT) technologies hold promise for enhancing device autonomy, particularly for energy-limited IoT systems. This paper presents experimental results on coherent and non-coherent transmit diversity approaches for WPT, tested in the near field using the Techtile testbed. We demonstrate that a fully synchronized beamfocusing system achieves a 14 dB gain over non-coherent transmission, consistent with the theoretical 14.9 dB gain for a 31-element array. Additionally, phase alignment errors below 20° result in less than 1 dB of gain loss, while errors exceeding 40° lead to losses over 3 dB. These findings suggest that phase coherency requirements for WPT can be relaxed, and that scaling the number of antennas is a promising strategy for improving power transfer efficiency.
\end{abstract}


\glsresetall
\section{Introduction}

\Acrfull{wpt} technologies hold a great potential to increase autonomy of devices in a very convenient way. They have attracted ample attention in view of the growing number of deployed energy-limited \gls{iot} devices. Ultimately, they could provide a theoretically infinite lifetime and enable energy-neutral operation of the \gls{iot} devices, i.e., the delivered energy fully covers the consumption needs of the devices. Many studies have addressed the main challenges related to getting sufficient energy across and increasing the efficiency of the \gls{wpt}, covering aspects such as, signals and systems or hardware solutions~\cite{Huang2019Wireless, Wagih2020Antennas, VanMulders2022, 7867826}. Large and distributed antenna systems open interesting new opportunities to focus the power in beams and spots~\cite{9743350, 8955833}. In particular, coherent approaches have been shown to theoretically provide spectacular improvements in transfer efficiency. However, these systems require a couple of tough problems to be resolved, including the acquisition of \gls{csi}~\cite{Deutschmann23ICC} and the synchronization of many distributed transmitters~\cite{10414364}, among others. To combat these challenges, non-coherent approaches are being proposed for multi-antenna systems. An example is employing phase sweeping transmit diversity~\cite{hiroike1992combined, Clerckx2018TransmitDiversity, VanM2410:Single}. In this method, the antennas do not require \gls{csi} and only transmit phase-shifted signals at a single carrier.
We have experimentally studied this technique and others in our testbed Techtile~\cite{CallebautTechtile} in the RF near-field\footnote{Note, in this work, the term \textit{near-field} refers specifically to the electric near-field region of the antenna array. While \textit{near-field} in the context of \gls{wpt} often implies electromagnetic coupling-based charging, this interpretation does not apply to our study.} and report on the promising results in this paper. Furthermore, We have performed experiments to validate the beamfocusing capabilities of distributed \gls{mimo} transmission for \gls{wpt} with different degrees of phase accuracies. The results confirm the theoretical beamfocusing gains for both non-coherent and coherent \gls{wpt} and confirm the hypothetical power spot size of $\lambda/2$. This raises the confidence level that, scaling up the number of antennas, is an effective strategy.  Moreover, the experiments show that phase alignment errors of up to \SI{20}{\degree}, still allow to achieve beamfocusing gains, with only a loss of \SI{1}{dB} with respect to the fully coherent case with a 31-element array, yet the losses rapidly increase beyond \SI{40}{\degree} errors.

This paper is further organized as follows. In the next section, we explain the experimental environment and conditions and in particular the approach to achieve synchronized transmission. The investigated \acrfull{wpt} strategies and experiments are introduced in~\cref{sec:meas} and~\cref{sec:setup} respectively. %
The measurements results and their evaluation are presented in~\cref{sec:eval-exp}. Finally,~\cref{sec:concl} summarizes the main conclusions of this study and provides an outlook for future work.

\begin{figure*}[h]
        \centering
        \begin{subfigure}{0.5\columnwidth}
        \centering
        \includegraphics[height=6cm]{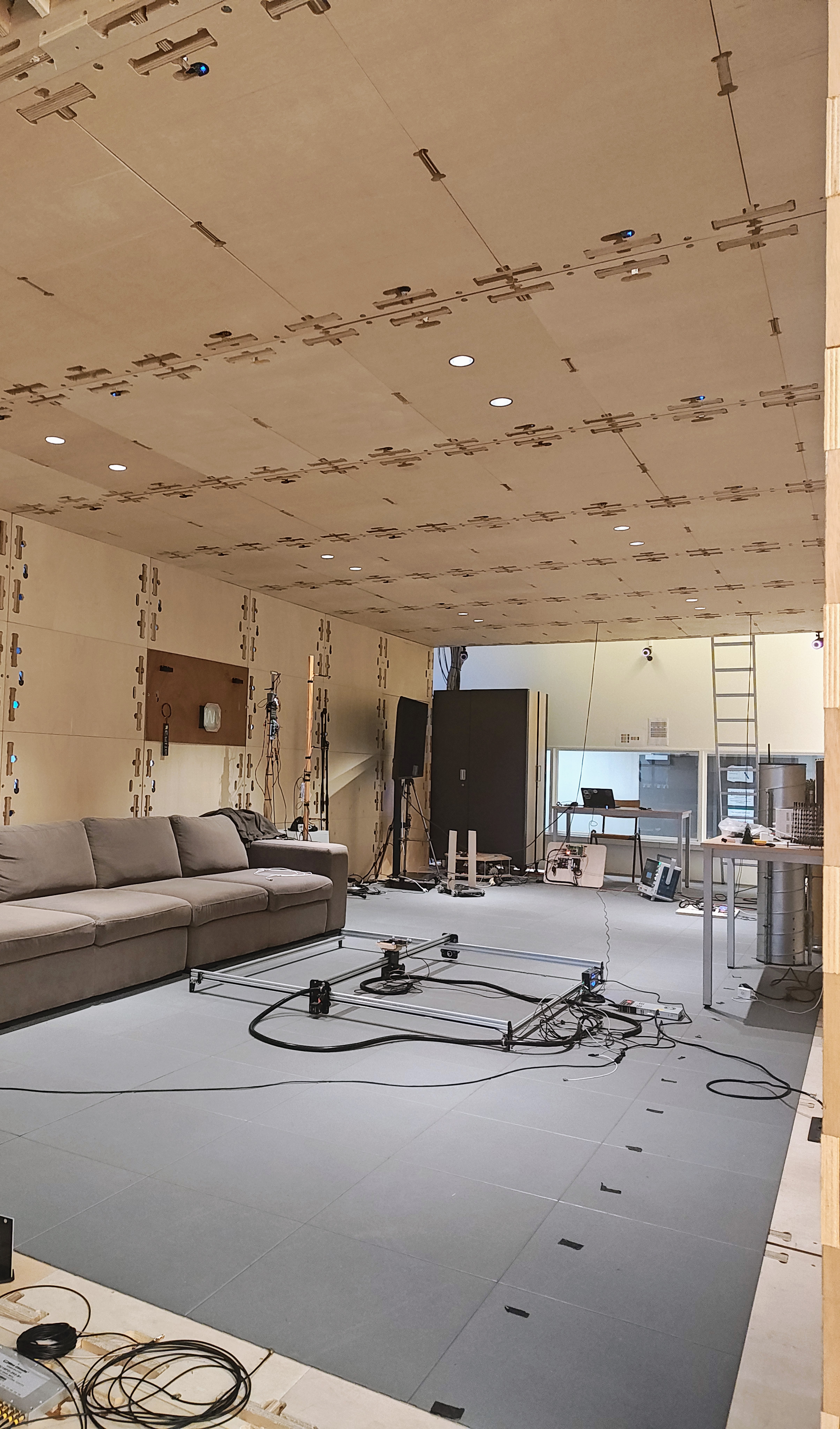}
        \end{subfigure}%
\begin{subfigure}{\columnwidth}
\centering
         \includegraphics[height=6cm]{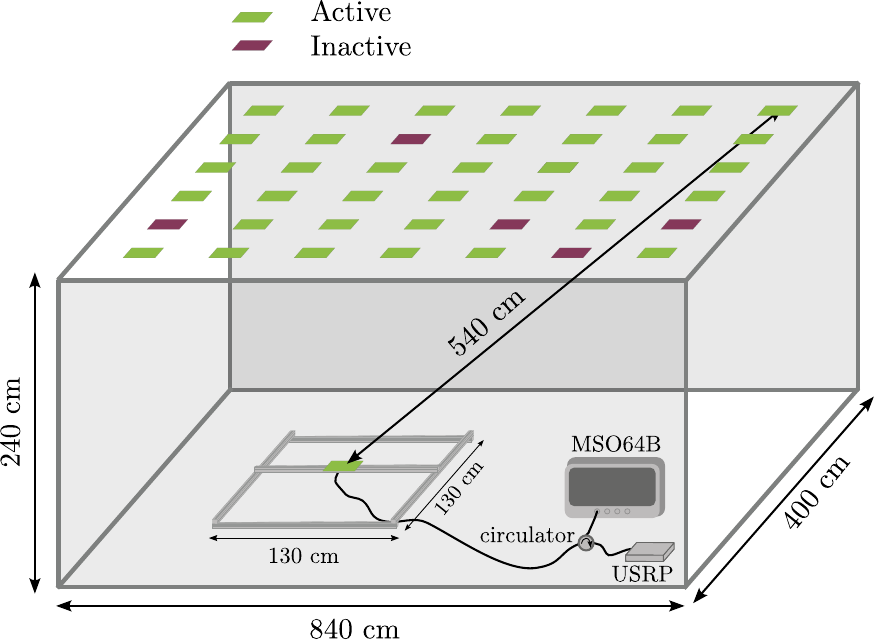}
\end{subfigure}
\caption{Picture and illustration of the experimental set-up Techtile~\cite{CallebautTechtile}, with transmitters on the ceiling and a movable antenna connected to the measurement equipment. The movable antenna allows to spatially sample the received power over a grid of \num{130} by \SI{130}{\centi\meter}.}
\label{fig:setup}
\end{figure*}

\begin{table*}[h]
    \centering
    \caption{Overview of \gls{wpt} strategies, their used acronyms in figures, the applied downlink transmit phase and the synchronisation requirement needed to provide \gls{rf} \gls{wpt}. }
    \label{tab:exp}
    \renewcommand{\arraystretch}{1.5}
    \begin{tabular}{l p{3.5cm} l p{7cm} c c c}
    \toprule
    & & & & \multicolumn{3}{c}{Sync. Requirements}\\
    Acr. & \gls{rf} \gls{wpt} Strategy & Applied TX phase & Short description & Time & Freq. & Phase\\
    \midrule
         \BF{} & beamfocusing (near-field) or beamforming (far-field) & $-\hat{\varphi}$ & In the beamfocusing strategy, reciprocity-based downlink \gls{wpt} is used, where the phase $\hat{\varphi}$ at all antennas is estimated using an uplink pilot. & \tikzcmark & \tikzcmark &  \tikzcmark \\
          \SISO{} & single antenna& N.A. & Only one antenna is used during transmission. & \tikzxmark & \tikzxmark & \tikzxmark \\
          \AS{} & Random-phase sweeping & $\varphi_{\text{AS}}\sim \mathcal{U}(0, 2\pi)$ & All transmitters change their phase every \SI{10}{\milli\second} at the same carrier frequency.& \tikzcmark & \tikzcmark & \tikzxmark \\
          \GBF{} & BF with Gaussian noise& $-\hat{\varphi}+\varphi_{\text{G-BF}}$ & After uplink training, each antenna adds a phase $\varphi_{\text{G-BF}}\sim \mathcal{N}(0, \sigma_\varphi^2)$ drawn from a normal distribution. This is used to study the effect on the accuracy of the phase synchronization.& \tikzcmark & \tikzcmark & \tikzpmmark \\ \bottomrule
        \end{tabular}
\end{table*}

\section{Techtile Experimental Environment and Synchronization Approach}

\subsection{Techtile Testbed}

The Techtile measurement infrastructure~\cite{CallebautTechtile} (\cref{fig:setup}) serves as a versatile, multi-functional testbed for experimental research on innovative communication, positioning, sensing, and federated learning technologies exploiting spatially distributed resources. Modularity is achieved by constructing the ceiling, floor, and walls out of 140 detachable tiles. Each tile is equipped with a custom \gls{poe} board with power supply up to \SI{90}{\watt}, a \gls{usrp} B210 with two antennas and a Raspberry~Pi~4 for edge processing. The backbone of Techtile consists of a central server, enabling communication, sub-microsecond time synchronization and power to all tiles over a single Ethernet connection. An in-depth review on the multi-functionality of Techtile can be found in \cite{CallebautTechtile}. \arxiv{Since the publication of \cite{CallebautTechtile}, the testbed has been extended with new hardware facilitating:
\begin{itemize}
    \item Sub-centimeter 3D positioning and tracking accuracy thanks to the Qualisys motion capture system.
    \item Accurate time and frequency synchronization through \gls{lo} distribution.
    \item Automatic 2D spatial sampling due to the OpenBuilds ACRO CNC system.
\end{itemize}}

\subsection{Synchronization Capabilities of Techtile}
This work examines various levels of synchronization, made possible by Techtile's capabilities in achieving a phase-coherent system. The methods used for time, frequency, and phase synchronization are outlined below.

\subsubsection{Time Synchronisation}
Time synchronization is achieved by distributing a \gls{pps} signal to all \glspl{usrp} using Octoclocks. Before starting a measurement, an absolute time is agreed upon by all \glspl{usrp}. To facilitate this, the \glspl{usrp} participating in the experiment connect to the same \gls{zmq} server. After establishing the connection, the server sends a start command, which is received by all \glspl{usrp} within a few milliseconds. Upon receiving the command, each \gls{usrp} waits for a \gls{pps} trigger, at which point the time is set to zero, thereby establishing a common absolute time frame. From this point forward, all \glspl{usrp} operate on the same time reference, and subsequent commands are executed at specific times to maintain time synchronization.

\subsubsection{Frequency Synchronisation}
Frequency synchronisation is achieved through the distribution of a \SI{10}{\mega\hertz} signal using Octoclocks to all \glspl{usrp}. This reference clock is used to synthesize the required clocks on the \gls{fpga} and transceiver.

\subsubsection{Phase Synchronisation}
The \gls{usrp} B210 devices have two separate \glspl{lo} to work in \gls{fdd} mode, i.e., transmitting and receiving simultaneous on two distinct frequencies. In the case of the \gls{usrp} B210, this results in a TX and RX \gls{lo} that are not phase-locked, i.e., the phase relation between the two \gls{lo} is unknown and nondeterministic.\footnote{To be complete, if the \glspl{sdr} are configured in \textit{integer \gls{pll} mode}, the phase relation between the \glspl{sdr} are deterministic due to the dividers present in the \gls{rf} chains.} 
However, this phase relation is required to perform coherent downlink beamforming. It relies on the fact that the transmit and receive frontends are reciprocal, i.e., the phase relation between them is known. In this setup, the phase differences (due to path lengths and electrical components) between the TX and RX frontends and \glspl{lo} are measured. More details regarding the reciprocity calibration can be found in~\faicon{github}\footnote{\url{github.com/techtile-by-dramco/NI-B210-Sync}}.

\section{Multi-Antenna \Gls{rf} \Gls{wpt} Strategies}\label{sec:meas}

In this work, we are studying the effect of the level of synchronization accuracy required to perform \gls{rf} \gls{wpt} in distributed systems. Four different strategies are considered and summarized in~\cref{tab:exp}. Each of those strategies requires a different synchronization level.
\begin{figure}[htbp]
    \centering
\begin{tikzpicture}
\begin{axis}[
width=0.7\columnwidth,
height=0.7\columnwidth,
colorbar,
colorbar style={ylabel={},ytick={-65,-60,...,-45},
ylabel={\footnotesize RSS in \SI{}{\dBm}}, 
xshift=-0.2cm,      %
width=0.4cm,        %
ytick pos=right,    %
ytick distance={5}, %
tick align=center,  %
},
colormap/viridis,
point meta max=-42.9938799584875,
point meta min=-65.8,
tick align=outside,
tick pos=left,
x grid style={darkgray176},
xmin=-0.5, xmax=299.5,
xtick style={color=black},
xtick={0,40,80,120,160,200,240,280},
xticklabels={0.0,0.5,1.0,1.5,2.0,2.5,3.0,3.5},
y grid style={darkgray176},
ymin=-0.5, ymax=299.5,
ytick style={color=black},
ytick={0,40,80,120,160,200,240,280},
yticklabels={0.0,0.5,1.0,1.5,2.0,2.5,3.0,3.5},
xlabel={Distance in wavelengths},
ylabel={Distance in wavelengths},
]
\addplot graphics [includegraphics cmd=\pgfimage,xmin=-0.5, xmax=299.5, ymin=-0.5, ymax=299.5] {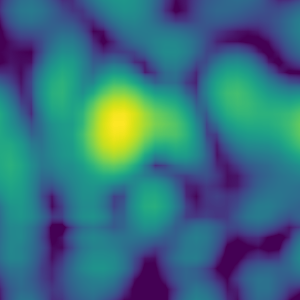};
\end{axis}

\end{tikzpicture}
    \caption{Heatmap of the \gls{rss} resulting from coherent beamfocusing (\BF{}), measured with the movable receive antenna over the full grid. The power spot is approximately $\lambda/2$ in diameter\arxiv{ (when considering a \SI{-3}{ \dB} power threshold)}.}
    \label{fig:heatmap}
\end{figure}

\subsection{\Acrlong{siso} (\SISO{})}
The \gls{siso} strategy is included in this work as a benchmark to compare the obtained gains of the other strategies. In the \SISO{} case, only one antenna is used to perform \gls{wpt}. Hence, no form of synchronization is required.

\subsection{Random-Phase Sweeping (\AS{})}
When the system is not phase-coherent, but provides time and frequency synchronization, random-phase sweeping can be used to illuminate dead spots.\footnote{Furthermore, this strategy can also be employed if the device is unable to send and uplink pilot and its location is unknown.} By randomly changing the phase of all antennas at regular intervals, fluctuations in the received power can be induced, yielding different levels of constructive interference at different time and spatial instances. This is especially interesting when the average received power is below the required threshold for a device to get charged. A peak in power could ``jump start'' the device. Multi-carrier systems are examples where the different carriers induce large fluctuations, yielding high \gls{papr} signals. However, in recent studies, single-carrier transmissions with adaptive phase-changes have shown to be more promising than multi-carrier systems for \glspl{end}~\cite{VanM2410:Single}.

When transmitting with the same power, the statistical expectation of the power gain with respect to \SISO{} is the sum of the received powers of all individual antennas~\cite{VanM2409:Keeping}.  This is sometimes also called the non-coherent gain. In literature, the gain is often expressed as $10\log_{10}(M)$, with $M$ the number of antennas. This is the case when the large-scale fading from all antennas to the device are the same. 
However, in near-field, the distributed antennas experiences different large-scale fading.

\subsection{Near-Field Beamfocusing (\BF{})}\label{sec:bf}
In a fully coherent system, the phase relations between all antennas are known. Hence, these systems need to be time, frequency, and phase synchronized. In the case when the devices are located in the near-field of the system, the term beamfocusing is used, while when operating in the far-field, beamsteering is used. In the latter case, under \gls{los} conditions and for a single user, the antenna array physically forms a beam directed toward the desired user. However, in the near-field, the received signals have a spherical wavefront, illuminating a specific spot rather than forming a beam, which is why the term beamfocusing is commonly used in the literature. Analogously, beamsteering acts like a flashlight, while beamfocusing resembles a spotlight. The term beamforming will be used as a general term to depict both beamsteering or beamfocusing, depending on the near- or far-field condition.

In our setup, beamfocusing is performed by estimating the phase $\hat{\varphi}$ of the received signal at all antennas, through a single-tone uplink signal transmitted by the device. This device will be called \gls{end} in the remainder of this manuscript. This indicates that it is the target device, where we want to transfer energy to. 
When performing downlink \gls{wpt}, the opposites of the estimated phases are applied when transmitting a single-carrier tone at all antennas, yielding constructive interference at the target location/\gls{end}. The corresponding ``spotlight'' will be called the power spot or focal region. This considered region is the area where the received power level is not lower than \SI{3}{dB} with respect to the target location. Such a power spot is visible in~\cref{fig:heatmap}. 

When transmitting with the same power, the expected gain with respect to a \SISO{}\footnote{In literature $10\log(M)$ is used, as they assume that with $M$ antennas the transmit power is also $M$ times lower, which is not considered in this work.} system is $20\log(M)$ and $10\log(M)$ compared to the \AS{} approach.

\subsection{Beamfocusing with phase errors (\GBF{})}\label{sec:GBF}
While \GBF{} is technically not a \gls{wpt} technique, the performance degradation with respect to the achievable phase-coherency case is studied by artificially applying a phase error to the estimated uplink phase. It allows evaluating the loss in gain with respect to the fully phase-coherent system. The applied transmit-phase during downlink \gls{wpt} is determined as,
\begin{equation*}
    -\hat{\varphi}+\varphi_{\text{G-BF}},
\end{equation*}
where $\hat{\varphi}$ is the estimate of the uplink pilot phase and $\varphi_{\text{G-BF}}\sim \mathcal{N}(0, \sigma_\varphi^2)$ is a random phase drawn from a normal distribution with standard deviation $\sigma_\varphi$.

When the standard deviation is high, this yields on average (in space and time) the same results as the \AS{}. However,\GBF{} would result in a constant received power, as the phase is constant during \gls{wpt}, which could be potentially a dead spot. While in \AS{}, due to the constantly changing phases, the same location would experience different power levels during \gls{wpt}.

\section{Experiments}
\label{sec:setup}
The aforementioned Techtile infrastructure is used to compare the proposed wireless power strategies in a real world environment. To confine the measurement size, the received power is measured in a 2D plane of \SI{130}{cm}$\times$\SI{130}{cm}, as sketched in~\cref{fig:setup}. At the infrastructure side, 1 or 31 of the distributed patch antennas in the ceiling transmit a \SI{920}{\mega\hertz} continuous sine wave with a transmit power of \SI{3}{\dBm}. Depending on the strategy, the transmitters are time and/or phase synchronized. For reciprocity-based downlink transmission, an uplink pilot is used to determine the received phase at each antenna, as detailed in~\cref{sec:bf}.

The antenna at the \gls{end} side is moved by the ACRO CNC machine in the delimited plane. The same patch antenna is used at the device and infrastructure side and are orientated towards each other. This enables repetitive received power measurements. As can be seen in \cref{fig:setup}, the \gls{end} antenna is connected via a circulator to the oscilloscope and \gls{end} \gls{usrp}. The circulator ensures that the transmitted signal from the \gls{usrp} is transmitted via the antenna, while the receive signals are relayed to the Tectronix MSO64B oscilloscope to measure the received power. The exact position and orientation of the receiver is obtained through the Qualisys system.  The processing scripts and data can be found on~\faicon{github}\footnote{\url{github.com/techtile-by-dramco/experiments/tree/main/02_reciprocity_based_WPT}}.

\begin{figure*}
    \centering
    \begin{tikzpicture}

\definecolor{darkgray176}{RGB}{176,176,176}
\definecolor{steelblue31119180}{RGB}{31,119,180}

\begin{axis}[
width=0.9\textwidth,
height=0.4\textwidth,
tick align=outside,
tick pos=left,
x grid style={darkgray176},
xlabel={Phase error std. dev. $\sigma_\varphi$ in ${}^\circ$},
xmin=-5, xmax=180,
extra y ticks={-45,-50,-55,-60,-65,-70,-75},
xtick style={color=black},
y grid style={darkgray176},
ylabel={Received power in dBm},
ymin=-77, ymax=-42,
ytick style={color=black},
grid style=dotted,
ymajorgrids,
xmajorgrids,
clip mode=individual,
legend style={
                at={(0,0)}, anchor=south west, %
                align=left, %
                scale=0.8 %
            },
            legend cell align={left},
]
        \addplot[
            only marks, %
            mark size=0.8pt, %
            forget plot,
            opacity=0.5,
            fill=G-BF,
            draw=G-BF
        ] table [x=x, y=y, col sep=comma] {figures/gaus-gain.dat};

        \addplot[
            forget plot,
            thick,
            G-BF
        ] table [x=x, y=y, col sep=comma] {figures/gaus-mean-gain.dat};
        
\path [draw=BF, semithick]
(axis cs:0,-45)
--(axis cs:1.8e+02,-45);

\node at (180,-45) [anchor=west, font=\tiny,BF] {BF};
\node at (0,-45) [anchor=south west, font=\tiny,BF] {-45};

\path [draw=AS, semithick]
(axis cs:0,-59)
--(axis cs:1.8e+02,-59);

\node at (180,-59) [anchor=west, font=\tiny,AS] {\AS{}};
\node at (0,-59) [anchor=south west, font=\tiny,AS] {-59};

\path [draw=SISO, semithick]
(axis cs:0,-66)
--(axis cs:1.8e+02,-66);

\node at (180,-66) [anchor=west, font=\tiny,SISO] {SISO};
\node at (0,-66) [anchor=south west, font=\tiny,SISO] {-66};

\node at (180,-55) [anchor=west, font=\tiny,G-BF] {G-BF};

\end{axis}

\end{tikzpicture}
    \caption{The \gls{rss} in the case of \SISO{}, \AS{}, and \BF{} is plotted for a single location). The phase synchronization accuracy (\GBF{}) is evaluated by adjusting the transmit phase as described in~\cref{sec:GBF}, each point represents one realization of the random phases and the line depicts the median (P50). Each realization yields a potential different receive power. \arxiv{The standard deviation of the phase error in degrees is on the x-axis.}}
    \label{fig:scatter}
\end{figure*}
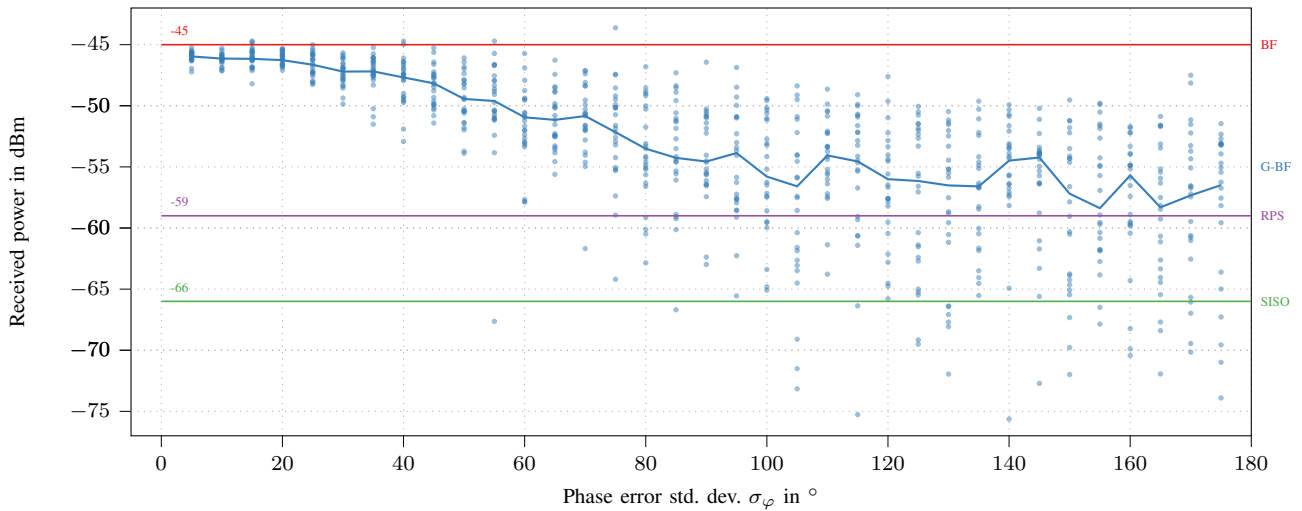

\begin{figure*}
    \centering
    \input{figures/power-cdf}
    \caption{\Gls{ecdf} of the measured received power using the movable \gls{end} antenna across the full grid for all studied strategies \gls{wpt}. The received powers outside and inside the focal region are plotted separately. The x-axis is limited to \SI{-80}{\dBm} to enhance readability, although power samples drop as low as \SI{-100}{\dBm} for the \BF{} strategy outside the focal region. The plot allows to examine i) the gain in the power spot region, ii) the power spread, and iii) the potential reduction in received power outside the target location in case of the \BF{} strategy.}
    \label{fig:cdf}
\end{figure*}
\section{Evaluation}
\label{sec:eval-exp}
The strategies are evaluated based on the received signals strength in the power spot and its size. Additionally, the effect of the synchronization level on the average power inside and outside the focal region is investigated.

\subsection{Power spot size}
As can be observed in~\cref{fig:heatmap}, \textbf{the created power spot (\BF{}) has a diameter of approximately $\lambda/2$ in diameter}.\arxiv{\footnote{The measured power spot area is \SI{0.03}{\square\meter}.}} This means, that outside of this area, the received power is lower than \SI{3}{dB} with respect to the target location. This is in line with the theoretical expectation of the power spot size~\cite{9048825}. 

\subsection{Effect of synchronization accuracy on the average received power}
The impact on the system synchronization capability/accuracy for the received power in the focal region can be seen in~\cref{fig:scatter}.  
\textbf{In case of a fully synchronized system (\BF{}), the average received power is \SI{-45}{dBm}. This is \SI{14}{dB} higher than the \AS{} case, which is in line with the expected gain of \SI{14.9}{\dB} (corresponding to} $10 \log_{10}(M)$\textbf{)}, as outlined in~\cref{sec:bf}. 
In turn, the \AS{} strategy has a gain of \SI{7}{dB} with respect to the \SISO{} case. Note that here the \SISO{} antenna is the one closest to the target location and thus has the largest path gain.  

With an increased phase error (\GBF{}), the received power is decreasing, moving closer to the \AS{} case, as can be seen when comparing the P50 of the \GBF{} to the \AS{} plot. The P50 of the \GBF{} allows to investigate the loss in gain when having a certain degree of phase error. As can be observed, \textbf{for low standard deviations of the phase error (below \SI{20}{\degree}), the gain loss is below \SI{1}{dB}.} As of  \SI{40}{\degree}, we have a gain loss higher then \SI{3}{dB}. \textbf{This observation is promising for \gls{wpt}, indicating that the phase-coherency requirements can be relaxed.} These observations are again in line with theoretical expectations~\cite{D2_3}. Notably, the same conclusions cannot be made for communication, which would require tighter synchronization. 

\subsection{Effect of synchronization accuracy on the distribution of received power}
In most cases, we want to have as much power as possible in the intended location and as low as possible outside. 
The distribution, and more specifically the \gls{cdf}, of the received powers for the different synchronization accuracies is shown in~\cref{fig:cdf}. The figure provides the received powers sampled in the full grid, as indicated in~\cref{fig:setup}.
This for both inside the focal region, i.e., an area of $\lambda/2$ by $\lambda/2$ around the target location, and outside the focal region. 

Observations:
\begin{enumerate}
    \item In the \SISO{} case, the received power is higher inside the focal region than outside. This is because the SISO antenna is positioned directly above the focal region. Hence, more power is expected right below the SISO transmit patch antenna.
    \item The \BF{} strategy results in the highest maximum and lowest minimum received power. ``Dark spots'' are created outside the focal region, having powers as low as \SI{-100}{dBm} (which is close to the measured noise floor). On the other hand, inside the focal region, the received powers are maximized, as also visible in~\cref{fig:heatmap}. 
    \item For \GBF{}, the small width of the \gls{cdf} indicates that the received powers within the focal region are very similar in the power spot area. This is a result of the phase noise added to the transmitted phases, which broadens the power spot, effectively smudging out the power spot. This effect is also observed and leveraged on in~\cite{Deutschmann23ICC}. 
    \item \AS{} exhibits a similar power distribution both inside and outside the BF region, which is expected since the \AS{} approach induces rapid power fluctuations that, when averaged over time, ideally occur uniformly across all positions. In some cases, higher power is observed in the BF region when using \AS{} compared to \GBF{} with a phase error standard deviation of $\pi/2$. This suggests that \textbf{\AS{} could be preferable in scenarios with low synchronization, where peak power is more desirable than average power}, as is sometimes the case for \glspl{end}~\cite{VanM2410:Single}.
\end{enumerate}

\glsresetall
\section{Conclusions and Future Outlook}\label{sec:concl}
This paper has reported on an experimental study of distributed beamforming for RF-based \gls{wpt}. In particular, a coherent approach is pursued. The results validate and demonstrate the great gains that can be achieved in efficiency as predicted in theory. The beam focusing capability is clearly present in the visualization of the measured receive power levels. An important condition to realize the power transfer efficiency gain is accurate synchronization of the transmitters, which poses a harsh challenge in practical deployments. Fortunately, the results indicate the promise of only small performance degradations even when having a standard deviation of the phase error as high as 40°, with only a \SI{3}{dB} loss. This would indicate that the update rate of the calibration and \gls{csi} estimation procedure could be lowered for \gls{rf} \gls{wpt}.
The current status is promising to enhance coverage and service levels for interaction with energy-neutral devices relying on RF-based \gls{wpt}. It calls for further R\&D efforts to progress low-complexity elegant solutions for achieving accurately synchronized transmitters. Furthermore, the scaling up towards more antennas and more devices to be serviced, poses additional challenges in terms of coordination, protocols and scheduling to be resolved.

\arxiv{\section*{Acknowledgements}
The authors would like to thank  Thomas Wilding from TU Graz for his valuable discussions on the subject, and Guus Leenders and Fran\c{c}ois Rottenberg from KU~Leuven for the help on getting Techtile synchronized.}

\printbibliography

\end{document}